\documentclass{emulateapj}
\usepackage{graphicx}
\usepackage{amsmath}
\usepackage{natbib}

\shorttitle{Retaining BHs with Recoil}
\shortauthors{Schnittman}

\begin{document}

\title{Retaining Black Holes with Very Large Recoil Velocities}
\author{Jeremy D.\ Schnittman}
\affil{Department of Physics, University of Maryland,
College Park, MD 20742}

\begin{abstract}
Recent numerical simulations of binary black hole mergers show the
possibility of producing very large recoil velocities ($> 3000$
km s$^{-1}$). Kicks of this magnitude should be sufficient to eject the final
black hole from virtually any galactic potential. This result has been
seen as a potential contradiction with observations of supermassive
black holes residing in the centers of most galaxies in the local
universe. Using an
extremely simplified merger tree model, we show that, even in the
limit of very large ejection probability, after a small number of
merger generations there should still be an
appreciable fraction $(>50\%)$ of galaxies with supermassive black
holes today. We go on to argue that the inclusion of more realistic
physics ingredients in the merger model should systematically {\it
  increase} this
retention fraction, helping to resolve a potential conflict between
theory and observation. Lastly, we develop a more realistic Monte
Carlo model to confirm the qualitative arguments and estimate
occupation fractions as a function of the central galactic velocity
dispersion. 
\end{abstract}


\maketitle

\section{INTRODUCTION}
\label{intro}

In the standard model of hierarchical structure formation,
the most over-dense regions of the universe collapse at early times to
form small gravitationally bound clumps of dark matter and
baryons. These proto-galaxies
subsequently merge with each other and form larger and larger objects,
up to the galaxies and clusters we see today
\citep{press74,efsta88,haehnelt93,cole00}. Observations of the
local universe suggest that most galaxies at redshift $z=0$ are also
host to a supermassive black hole (SMBHs)
near their centers \citep{richstone98}. Observations of distant
quasars in the Sloan survey suggest that
these SMBHs are common up to redshifts of
at least $z\sim 6$ \citep{fan01}. One natural explanation is that the
SMBHs originally formed
in the early universe and then simply followed their host galaxies
throughout multiple generations of hierarchical mergers: whenever a
pair of galaxies merged, their central black holes would sink together
through dynamical friction and eventually merge themselves, resulting
in a single, more massive galaxy with a single, more massive BH at its
center \citep{kauffmann00,cattaneo01}. 

In the absence of special symmetries, the final black hole will
receive a linear momentum recoil from the gravitational radiation
emitted during the binary merger process
\citep{bekenstein73,fitchett83}. Recent numerical relativity
simulations suggest that this recoil cannot exceed $175$ km
s$^{-1}$ for non-spinning black holes \citep{herrm06,baker06,gonza06},
but may be as large as $\sim 3600$ km s$^{-1}$ for maximally spinning
equal-mass black holes with spins anti-aligned in the orbital plane
\citep{gonza07,campa07}. In this latter case, the resulting BH would
almost certainly be ejected from the gravitational well of its host
galaxy \citep{menou01,merri04}.

However, since observations of the local universe suggest that SMBHs
are quite common, there appears to be something
preventing these ejections. As a first attempt at resolving
this conflict, \citet{schnittman07} carried out Monte Carlo
simulations of the binary BH recoil for a large range of mass ratios
and spin orientations, finding a relatively small fraction $f_{1000}
\simeq 0.05-0.1$ of systems with kick velocities $\ge 1000$ km
s$^{-1}$. These fractions correspond to black holes with dimensionless
spin parameters of $a/M=0.9$, which appear to be typical from AGN
observations \citep{yu02,elvis02,wang06}. More recently,
\citet{bogdanovic07} proposed a
mechanism in which a circumbinary gas accretion disk would align the
spins of the two black holes with the orbital axis of the disk prior
to merger, thereby limiting the maximum recoil to $\lesssim 200$ km
s$^{-1}$ \citep{baker07}.

In this paper, we take a rather different approach to the problem and
question the very premise of the apparent conflict: Does a large
recoil velocity necessarily imply a small occupation fraction?
Furthermore, does a small occupation fraction at high redshift imply
that only a
small number of SMBHs will survive until today? If, for
example, $95\%$ of all BH mergers result in an ejection from the host
galaxy, does this really mean that at least $95\%$ of galaxies today
should be devoid of SMBHs? Using a binary tree merger model for
the hierarchical growth of BHs and their host galaxies, we show that
the answer to these questions is a resounding ``NO!'' Our simple
analytic predictions appear to agree well with recent work by \citet{volon07},
which employs a much more detailed merger evolution, including BH
evolution due to accretion, but also finds a large occupation fraction
at low redshift.

The formal mathematical argument is given below in Section
\ref{merger_model}, but a simple qualitative reasoning is this:
in every generation of the merger tree, the {\it number} of galaxies
decreases, so the {\it fraction} with BHs can easily increase. If every BH host
galaxy merges with an empty galaxy, then the fraction of galaxies with
SMBHs can double from $0.5$ to $1$. Including the losses due to ejected
BHs, we find a steady state solution with occupation
fraction $f = 1/(1+p_{\rm ej})$, where $p_{\rm ej}$ is the probability
that a merging BH is ejected due to gravitational recoil. 
In Section
\ref{physics_components}, we discuss a number of potential physics
ingredients that could be added to the simple model, arguing that each
one would systematically increase the fraction of observed SMBHs
today. In Section \ref{astro_model} we attempt to confirm some of
these arguments with a more astrophysically realistic Monte Carlo
merger model, and in Section \ref{discussion} we present our conclusions.

\section{SIMPLE MERGER TREE MODEL}
\label{merger_model}

\begin{figure}
\begin{center}
  \scalebox{1.0}{\includegraphics{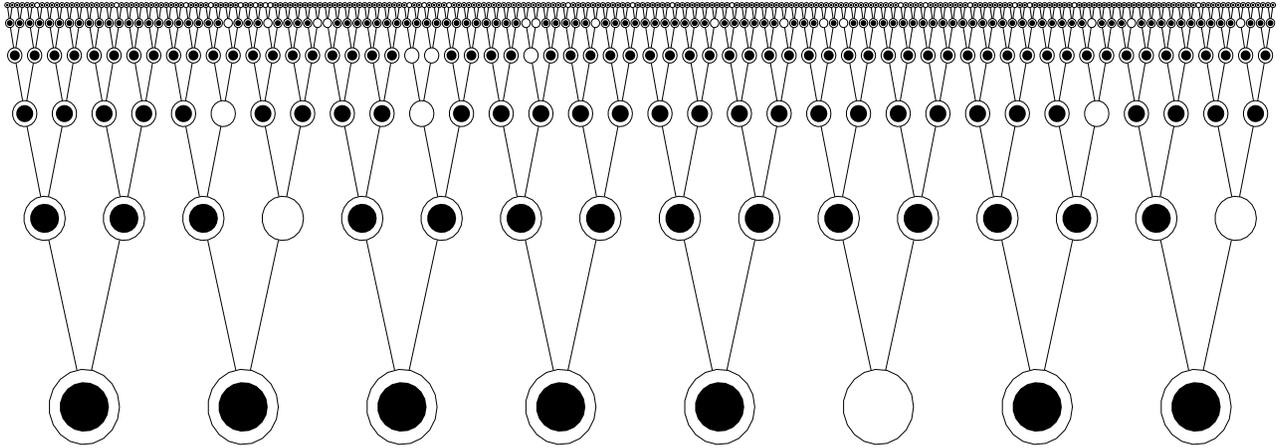}}
  \caption{\label{tree_01} Schematic of a binary merger tree for an
    ejection probability of $p_{\rm ej}=0.1$ after six
    generations. The circles represent 
    galaxies while the black dots in their centers symbolize the
    presence of a SMBH.}
\end{center}
\end{figure}

\begin{figure}
\begin{center}
  \scalebox{1.0}{\includegraphics{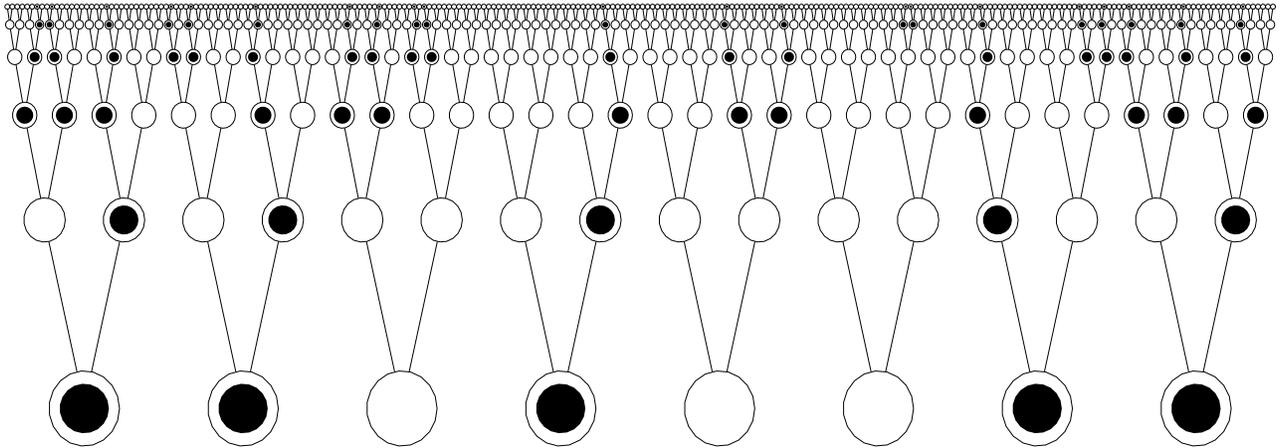}}
  \caption{\label{tree_09} Merger tree for an ejection probability of
  $p_{\rm ej}=0.9$ after six generations.}
\end{center}
\end{figure}

We begin with the simplest possible model for a galactic merger tree:
a binary tree where in each generation every galaxy merges with one
``spouse'' and produces a single ``child'' (see Figs.~\ref{tree_01}
and \ref{tree_09} for a schematic). Each parent galaxy may or
may not have a SMBH at its center, and if they both do, then the
respective black holes are assumed to merge into a single black hole,
which is then ejected due to gravitational recoil with a probability
of $p_{\rm ej}$. In the zeroth generation, before any mergers,
every galaxy is assumed to have a central black hole present. Thus in
the first generation, a fraction $f_1=1-p_{\rm ej}$ of galaxies will
have SMBHs. 

For a galaxy in the $(i+1)^{\rm st}$ generation, the probability of
having a black hole is 
\begin{eqnarray}\label{f_ip1}
f_{i+1} &=& 0\times(1-f_i)^2 \nonumber\\
& & + f_i(1-f_i) + (1-f_i)f_i \nonumber\\
& & + (1-p_{\rm ej})f_i^2 \nonumber\\
&=& f_i(2-f_i-p_{\rm ej}f_i),
\end{eqnarray}
where the first line corresponds to having two parents without
black holes, the second line represents either a black hole mother or
father, and the third line is the probability of both parents having
black holes, but the child black hole {\it not} getting ejected in the
merger. Taking the convergence limit of $f_{i+1}=f_i$ for large $i$,
we find
\begin{equation}\label{f_infty}
f_{\infty} = \frac{1}{1+p_{\rm ej}}.
\end{equation}

This remarkable result suggests that, in the limit of infinite recoil
velocity, $50\%$ of all galaxies will still retain a central black
hole. This is possible because, during one generation of galaxy
mergers with $f_i=0.5$, while effectively half of the black holes are
removed by the recoil, the total {\it number} of galaxies is also cut
in half, thus maintaining a constant {\it fraction} of SMBHs.

In Figures \ref{tree_01} and \ref{tree_09} we show examples of the
binary merger trees that obey the simple rules described above. The
heavy black circles represent black holes, while the empty circles are
galaxies without SMBHs. For
ejection fractions $p_{\rm ej}=0.1$ (Fig.\ \ref{tree_01}) and $p_{\rm
  ej}=0.9$ (Fig.\ \ref{tree_09}) we show six generations of galaxy
mergers, beginning with $f_1=1-p_{\rm ej}$. In both cases, the system
approaches the equilibrium fraction $f_{\infty}$ of equation
(\ref{f_infty}) after only a few generations. 

For a more quantitative estimate of the ``convergence time'' that it
takes to reach equilibrium, we see from equation
(\ref{f_ip1}) that
\begin{eqnarray}
f_1 &=& 1-p_{\rm ej}, \nonumber\\
f_2 &=& 1-p_{\rm ej}+p_{\rm ej}^2-p_{\rm ej}^3, \nonumber\\
f_3 &=& 1-p_{\rm ej}+p_{\rm ej}^2-p_{\rm ej}^3
+p_{\rm ej}^4-p_{\rm ej}^5+p_{\rm ej}^6-p_{\rm ej}^7, \nonumber\\
&\vdots& \nonumber\\
f_i &=& \sum_{j=0}^{2^i-1} (-1)^j \, p_{\rm ej}^j \, .
\end{eqnarray}
This series not only converges to equation (\ref{f_infty}) as
expected, but does so at quite a rapid pace. Even for extremely
large ejection rates of $p_{\rm ej}=0.95$ (i.e.\ $f_1=0.05$), $f_i$ is
within $5\%$ of its asymptotic value of $f_\infty=0.513$ after just
five generations. Because of this rapid convergence, we find that the
final results are independent of the initial occupation fraction
$f_0$.

\section{ADDITIONAL PHYSICS COMPONENTS}
\label{physics_components}

The model as described above is based on two major simplifying
assumptions: a perfect binary merger tree, and a constant ejection
probability. Of course, in reality both of these assumptions are in
all likelihood invalid. Here we list a number of possible
modifications based on more realistic astrophysics, and for each one
argue that their inclusion will only increase the fraction of SMBHs
observed today. 

The binary merger tree assumes that for a given generation, every
galaxy has exactly the same mass and merges with exactly one other
galaxy, halving the total number of galaxies in each subsequent
generation. In practice, most hierarchical merger simulations go {\it
  backwards} in time, using the extended Press-Schechter formalism
\citep{press74} to
estimate the progenitor masses at a given redshift \citep{volonteri03}. Many of these
simulations find that a single trunk of the merger tree dominates,
with a large number of small branches joining in at different
redshifts \citep{malbon06}. This means that the mass ratio for a typical
(proto-)galaxy merger can be significantly different than
unity. Assuming that the SMBH masses scale with their host masses,
this suggests that BH mass ratio will not be unity
\citep{sesana07}. While the Fitchett 
scaling for non-spinning black holes favors a mass ratio of about
three-to-one for maximum recoil \citep{fitchett83}, numerical
simulations imply that the 
super-massive kicks of $\gtrsim 1000$ km s$^{-1}$ require nearly-equal
masses and spins \citep{herrm07,gonza07,campa07,tichy07}, so a wider range of
mass ratios will tend to decrease the ejection probability $p_{\rm ej}$. 

Furthermore, as the galaxies evolve along the hierarchical merger
tree, their masses will increase, and thus so will their escape
velocities. Of course, the BHs are also growing in mass through
mergers and accretion during this time, but the recoil problem is
strictly scale-invariant: the kick velocity is a function only of the
mass {\it ratio} and the dimensionless spin parameters. Thus if the
masses of the BHs and their host bulges all double, the recoil will be
the same, but the escape velocity will be larger, and thus $p_{\rm
  ej}$ will be smaller. In this scenario, even if most of the BHs are
ejected at large redshifts, after a few generations $f_i$ will
grow rapidly due to a large number of ``single-parent'' mergers,
ultimately converging to a larger population fraction based on the
smaller value of $p_{\rm ej}$ at late times. 

Another effect that may increase the escape velocity is the formation
of a triple-BH system by the relatively prompt merger of three haloes
before their respective BHs have time to merge \citep{hoffman07}. In
many of these cases, one of BHs is ejected by Newtonian three-body
interactions, leaving a ``normal'' binary BH in a more massive host
galaxy (i.e.\ made of three parents instead of two),
further increasing the escape velocity and lowering the ejection
probability. However, this same three-body interaction may be strong
enough to give both the single and binary a large enough {\it
  Newtonian} recoil so that all three BHs are ejected.

In addition to these modifications to the merger tree physics and
escape velocities, there
are also a number of processes that more directly affect the actual
recoil velocity. Since the largest kicks are found in systems with the
BH spins in the orbital plane, any systematic effect that tends to
avoid this orientation will thus reduce the expected kick
velocity. One particularly strong influence on the spin orientation is
the torque produced by a single circumbinary accretion disk, which can
align both black hole spins with the orbital angular momentum
with high efficiency \citep{bogdanovic07}. In this orientation, the
maximum recoil should not be more than $\sim 200$ km s$^{-1}$
\citep{baker07}, well
below the escape velocities of most present-day galaxies. In the event
that there is no surrounding accretion disk (i.e.\ a ``dry merger''),
the two black hole spins may become aligned via spin-orbit resonances
\citep{schnittman04}, but this process likely requires somewhat
special initial conditions.

Lastly, there is also the possibility of creating a new SMBH 
{\it ex nihilo} during the galactic mergers, which typically are
accompanied by massive gas inflows to the center of the resulting
galaxy \citet{mihos94}. This rapid increase in
gas density will trigger a burst of massive star formation, which may
then proceed to form super-massive stars through runaway mergers
\citep{gurkan06}, in turn collapsing to form the seeds of
SMBHs, which will also be surrounded by copious amounts of
fuel to accrete more mass. In this way,
orphan black holes can appear in the merger tree, further increasing the
overall occupation fraction.

\section{ASTROPHYSICAL MERGER TREE}
\label{astro_model}

In an attempt to verify some of these qualitative claims, we now
develop a slightly more physical Monte Carlo merger model, motivated by basic
astrophysical arguments and observations of local SMBHs. Instead of
using a constant ejection probability and a perfectly binary merger
tree of equal-mass BHs and galaxies, we now consider an initially flat BH
mass distribution function $dN/dM_\bullet \sim M_\bullet^{-1}$, and
assume the galaxy mass is proportional to BH mass. If an occupied galaxy
merges with an empty galaxy, we adjust the final BH mass to ``agree''
with the total mass of both parent galaxies (physically represented by
some accretion episode). For an isothermal
sphere model, it turns out the galaxy mass is not important in
determining the escape velocity, which depends only on the velocity
dispersion: $v_{\rm esc} = 2\sigma_v$. The velocity dispersion is in
turn determined from the BH mass via the $M-\sigma$ relation of
\citet{ferra00,merri01}: 
\begin{equation}
M_\bullet \approx 1.3\times 10^8 \left(\frac{\sigma_v}{200 \mbox{ km
      s}^{-1}}\right)^{4.7} M_\odot \, .
\end{equation}

In each generation, a fraction $f_m$ of all the galaxies (selected
randomly) merge. In the limit of $f_m \to 1$ (as in the binary merger
tree), the low-mass tail of the distribution is prematurely depleted,
and if $f_m \to 0$, no evolution takes place at all. However, we find
that for the intermediate range $0.25 \lesssim f_m \lesssim 0.75$, the
net results are largely independent of $f_m$. For each merger, we
determine the final occupation as in the binary model: two empty
parents create an empty child, a single BH mother or father
will create a BH child, and in the case of two BH
parents, the child black hole will be ejected if the kick velocity is
greater than the escape velocity of the child galaxy. 

To determine the kick velocity in this more astrophysically realistic
model, we employ the analytic fits presented in \citet{schnittman07},
assuming all BHs are rapidly spinning with $a/M=0.9$. In this case,
for a given mass ratio $q\equiv m1/m2 \le 1$, $90\%$ of the mergers
should produce kick velocities less than $v_{90}$:
\begin{equation}
v_{90}(q) \approx \frac{17900 q^2}{(1+q)^5} \sqrt{(1-q)^2+1.4(1+q)^2}
 \mbox{ km s}^{-1},
\end{equation}
with the actual recoil $v_{\rm kick}$ selected randomly from the
cumulative distribution function \citep{schnittman07}
\begin{equation}
P_{\rm cdf}(v_{\rm kick}; q) = 10^{-v_{\rm kick}^2/v_{90}^2(q)} \, .
\end{equation}

\begin{figure}
\begin{center}
  \scalebox{0.8}{\includegraphics{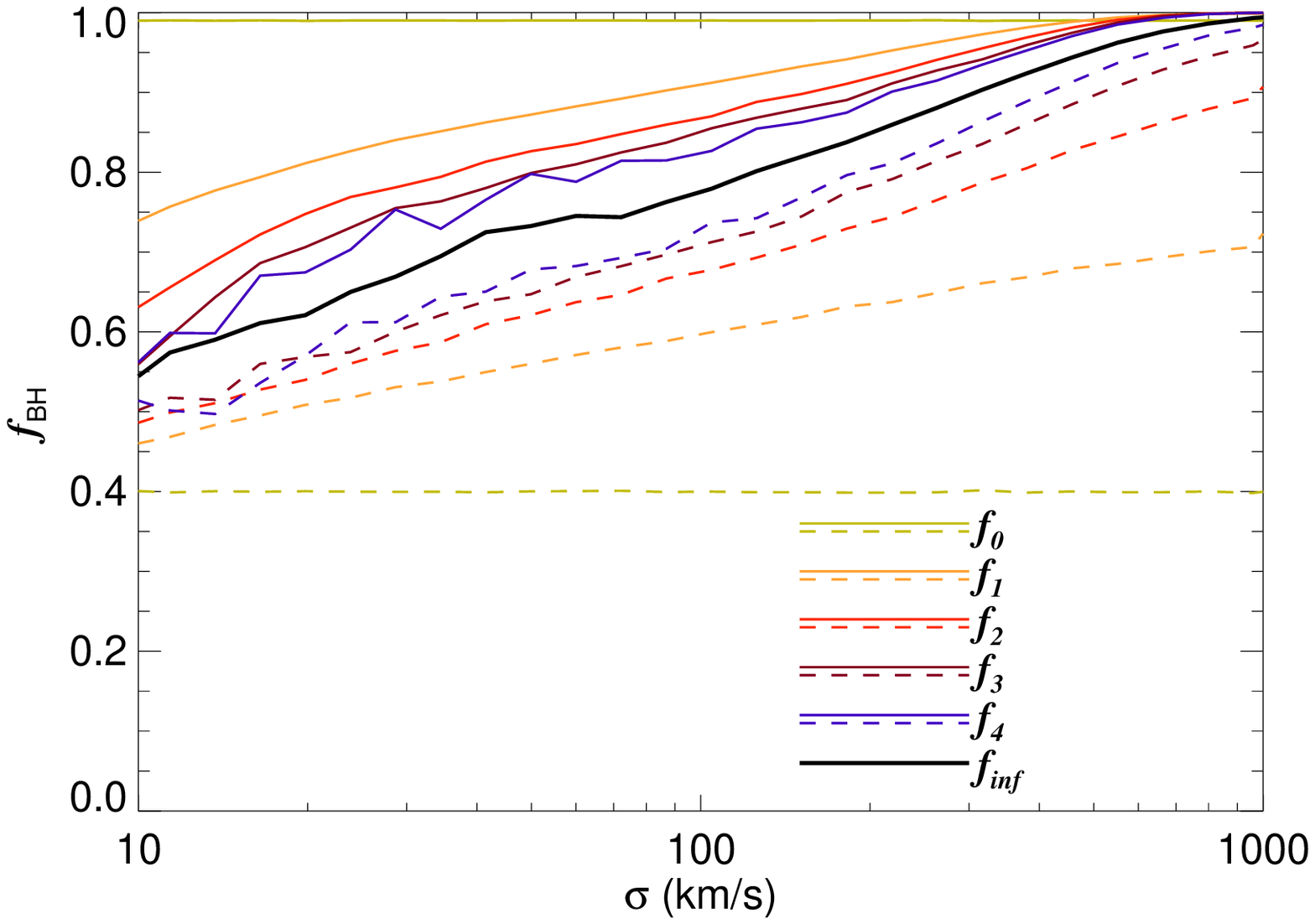}}
  \scalebox{0.8}{\includegraphics{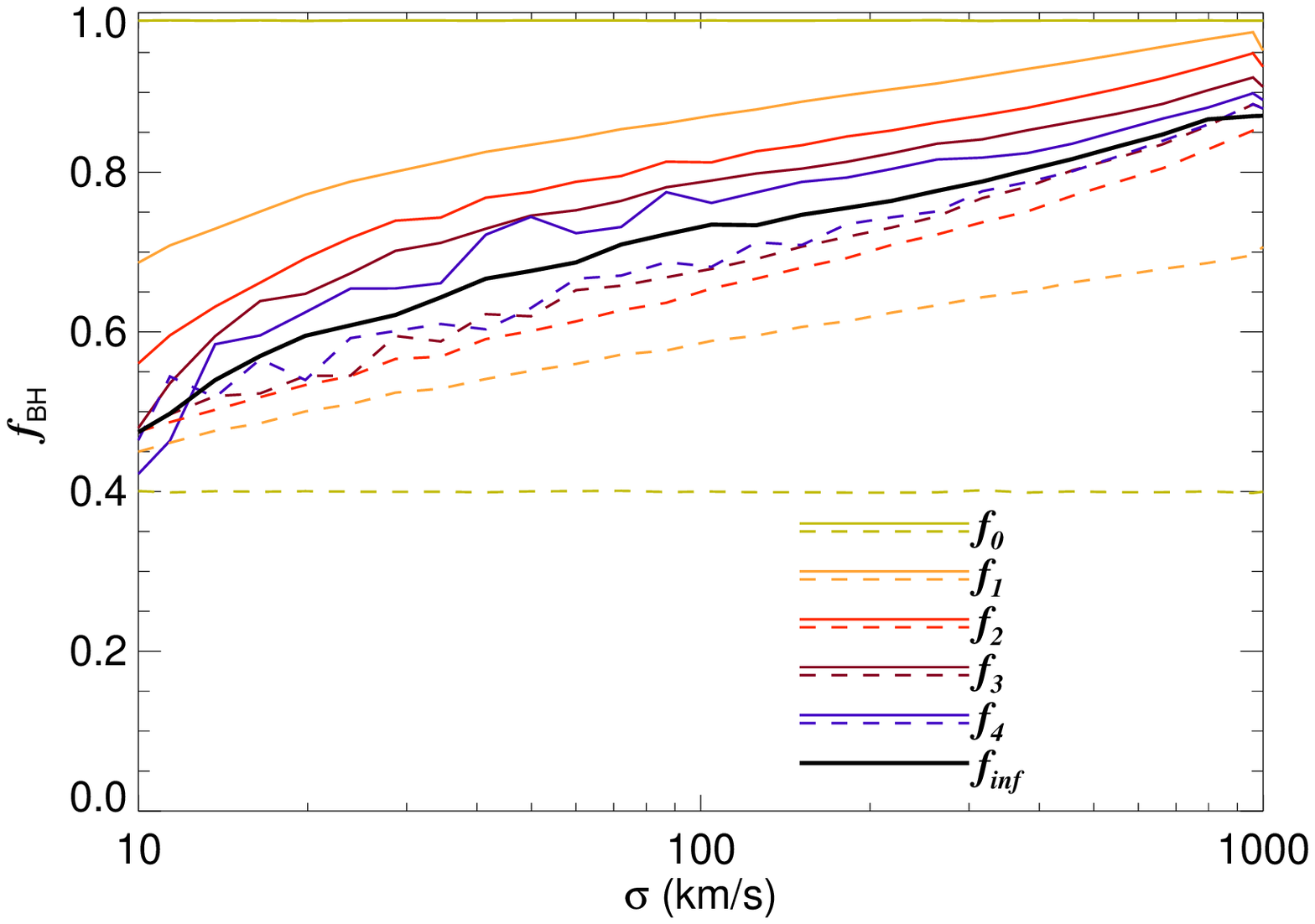}}
  \caption{\label{f_sigma} BH occupation fraction as a function of the
    galactic velocity dispersion $\sigma_v$ for the Monte Carlo merger model
    with $v_{\rm esc} = 2\sigma_v$ and the recoil velocity distribution of
    \citet{schnittman07} ({\it upper panel}). The solid curves
    correspond to an initial occupation fraction $f_0=0.99$ and the
    dashed curves $f_0=0.4$, with $f_i$ the occupation fraction after
    an average of $i$ mergers per galaxy. Both models converge to a single value of
    $f_{\infty}$, shown as a thick black curve. In the {\it lower
      panel}, we show the occupation fraction for an average kick
    velocity three times as large as that predicted by
    \citet{schnittman07}. In all cases, we set the merger fraction in
    each generation as $f_m=0.66$.}
\end{center}
\end{figure}

In the upper panel of Figure \ref{f_sigma}, we plot the occupation
fraction as a function of $\sigma_v$, employing the model
described above for two cases: an initial occupation of $f_0=0.99$
(solid curves) and for $f_0=0.4$ (dashed curves). In both plots, the
fractions $f_i$ refer to the average number of mergers for each
galaxy, {\it not} the number of generations ($N_{\rm mergers} \approx
N_{\rm gen} f_m/(2-f_m)$).  After a few
mergers, both cases converge to the steady-state case of $f_\infty$
(heavy black curve), just as in the simple binary merger model. In the
lower panel, we repeat the same calculation, now multiplying $v_{90}$
by a factor of three, to see the effect of extremely large
kicks. While the occupation fractions clearly decrease somewhat, they
remain significantly above $50\%$, just as predicted by the
binary merger model. 

We do not claim that this improved model should be seen as reliable
for making quantitative astrophysical predictions, but rather present it as
an application of the simple binary tree model and a confirmation of many of
the qualitative arguments from Section \ref{physics_components}. At
the same time, it is noteworthy that the occupation fractions for the
standard model (upper panel of Fig.~\ref{f_sigma}) seem to agree quite
closely with the much more detailed model of \citet{volon07}.
We anticipate that future observations should be able to measure this
distribution at larger redshifts and smaller $\sigma_v$, ultimately
explaining how the BH occupation fraction evolves in time. 

\section{CONCLUSIONS}
\label{discussion}

Based on a very simple binary merger tree model, we find that the
fraction of galaxies hosting a SMBH should be $\gtrsim 50\%$ even in
the limiting case of very large recoil velocities (or alternatively,
very small escape velocities). This would be particularly important at
high redshifts, when typical galaxies have small escape velocities and
the seeds of today's SMBHs are presumably formed. Including
qualitative arguments about
the hierarchical growth of the host galaxies, the ejection probability
$p_{\rm ej}$ will tend to decrease with cosmological redshift, further
enhancing the fraction of central BHs observed today. Including
spin-alignment effects from accretion will increase this fraction even
more. 

In Section \ref{astro_model}, we presented a somewhat more physical
Monte Carlo model and were able both to confirm the predictions of the
binary merger tree and also reproduce the basic results of the
detailed calculations of \citet{volon07}. 
We thus conclude that the very large kicks predicted by
numerical relativity should not necessarily be seen as contradictory to the
seemingly ubiquitous population of galaxies with SMBHs observed today.

\vspace{0.25cm}\noindent The author would like to thank Alessandra
Buonanno, Julian Krolik, and Cole Miller for helpful discussions and
comments. Partial support comes from NSF grant PHY-0603762. 


\clearpage

\end{document}